\documentclass[
prl,
twocolumn,
dvips,
aps]{revtex4}
\usepackage{amsmath,amssymb,amsfonts,graphicx}

\begin{document}

\title{Khinchin theorem and anomalous diffusion}

\author{Luciano C. Lapas$^{1,2}$, Rafael Morgado$^3$, Mendeli H. Vainstein$^{1,2}$, J. Miguel Rub\'{\i}$^2$ and Fernando A. Oliveira$^1$}
\email{fernando.oliveira@cnpq.br}
\affiliation{
$^{1}$Instituto de F\'{\i}sica and Centro Internacional de F\'{\i}sica da Mat\'{e}ria Condensada, Universidade de Bras\'{\i}lia, Caixa Postal 04513, 70919-970 Bras\'{\i}lia, Distrito Federal, Brazil\\
$^{2}$Departament de F\'{\i}sica Fonamental, Facultat de F\'{\i}sica, Universitat de Barcelona, Av. Diagonal 647, 08028 Barcelona, Spain\\
$^{3}$ Faculdade UnB Gama, Universidade de Bras\'{\i}lia, 72405-610 Bras\'{\i}lia, Distrito Federal, Brazil
}

\begin{abstract}
A recent paper~[M.~H.~Lee, {\em Phys. Rev. Lett.} \textbf{98}, 190601 (2007)] has called attention to the fact that irreversibility is a broader concept than ergodicity, and that therefore the Khinchin theorem [A.~I.~Khinchin, {\em Mathematical Foundations of Statistical Mechanics} (Dover, New York) 1949] may fail in some systems. In this Letter we show that for all ranges of normal and anomalous diffusion described by a Generalized Langevin Equation the Khinchin theorem holds.
\end{abstract}



\maketitle
\emph{1. Introduction.}---The ergodic hypothesis (EH) states that the time and ensemble averages of phase variables exist and are equal for  a stationary system. This hypothesis enables us to calculate thermodynamic quantities from the equations of motion of the particles and is crucial for the proof of basic  theorems in statistical mechanics~\cite{Lee07a,Khinchin49,Lee01,Boltzmann74,Costa03}. One of those theorems is the Khinchin theorem (KT)~\cite{Khinchin49}, of great importance, since it relates the ergodicity of a variable $p$ to the irreversibility of its autocorrelation function. In a recent work~\cite{Lee07a} based on the well established method of recurrence relations~\cite{Lee83}, it was asserted that, contrary to the KT, irreversibility is not a sufficient condition for ergodicity; in other words, the KT may not be valid for all systems. This demonstration poses a new challenge, \emph{i.e.}, that of determining in which systems the KT is valid.

Most of the experimental situations in which the EH does not hold arise in complex nonlinear or far from equilibrium structures where detailed balance is not fulfilled. A few examples are found in supercooled liquids~\cite{Ediger96,Kauzmann48}, glasses~\cite{Ediger96,Parisi97,Ricci00} and blinking nanocrystals~\cite{Margolin05}. The majority of those systems, however, apparently do not have an easy analytical solution, with an exception having recently been reported in~\cite{Hentschel07}. On the other hand, even anomalous diffusion treated with the Generalized Langevin Equation (GLE) can present closed solutions for the main expectation values, and can be used as a simple laboratory for the discussion of those properties. As we shall see, in this case, we can give a full description for the validity of the KT, even when the EH breaks down. Although anomalous diffusive processes are present also in the context of deterministic Hamiltonian maps~\cite{Shlesinger93}, we shall focus our attention on the GLE formalism.

In this Letter we work with stochastic processes and show that the KT is valid for all ranges of anomalous diffusion described by a GLE, even if this condition fails for Hermitian systems~\cite{Lee07a}.

\emph{2. Khinchin's theorem.}---Let $p$ be a dynamical stochastic variable (e.g., momentum operator) of a classical particle. The relaxation function can be written as 
\begin{equation}
R(t)=\frac{C_{p}(t)}{C_{p}(0)}\text{,}\label{R(t)}
\end{equation}
where the autocorrelation function of $p$ is given by $C_{p}(t)=\langle p(t)p(0)\rangle-\langle p(t)\rangle\langle p(0)\rangle$, and $\langle\ldots\rangle$ stands for an ensemble average. Explicitly, the KT states that if
\begin{equation}
R(t\rightarrow\infty)=0\text{,}\label{R(t->infty)}
\end{equation}
then $p$ is ergodic~\cite{Khinchin49}. In other words, the irreversibility condition is a necessary and sufficient condition for the validity of the EH. In Ref.~\cite{Lee07a}, it was claimed that for a system to be ergodic, it is necessary that  $0<W<\infty$, where
\begin{equation}
W=\int_{0}^{\infty}R(t')dt'\text{.}
\label{eq_W}
\end{equation}
If this condition is true and if $p$ refers to a Hermitian system, then irreversibility is not a sufficient condition for ergodicity, since the validity of Eq.~(\ref{R(t->infty)}) does not imply that $W<\infty$.

\emph{3. Diffusion phenomena.}---Diffusive dynamics is usually analyzed using the mean square displacement of the particles, which behaves in general as
\[
\langle \left[  x(t)-\langle x(t) \rangle \right] ^{2} \rangle \propto t^{\alpha} \text{,}
\]
 where the exponent $\alpha$ classifies the different types of diffusion: subdiffusion for $0<\alpha<1$, normal diffusion for $\alpha=1$, and superdiffusion for $1<\alpha\leq2$; for $\alpha=2$ the process is called ballistic~\cite{Morgado02, Costa03, Vainstein06a, Bao06, Lapas07}. According to Kubo's linear response theory~\cite{Kubo66}, the diffusion constant is given by
\begin{equation}
D=\lim_{t\rightarrow\infty}\frac{1}{2t}\langle \left[  x(t)-\langle x(t) \rangle \right] ^{2} \rangle=\frac{C_{p}(0)}{m^2}\int_{0}^{\infty}R(t')dt'\text{,}\label{D}
\end{equation}
where $m$ is the mass of the particle.
Thus, for normal diffusion $0<D<\infty$, for subdiffusion $D=0$, and for superdiffusion $D=\infty$. According to the result of~\cite{Lee07a}, ergodicity is only ensured for normal diffusion. Since at the time the KT was formulated most of the processes studied displayed normal diffusion with exponential relaxation, it is quite natural to inquire if the KT will be affected, for example, in the slow relaxation dynamics that occurs in anomalous diffusion. In the anomalous regime, power laws, stretched exponentials, and Bessel functions are only a few examples of the vast functional behavior that is possible for relaxation~\cite{Metzler04,Vainstein06a}.

A general description of the diffusion dynamics can be given by means of the GLE, which was developed by Mori~\cite{Mori65} using a Hermitian formulation, allowing to describe all diffusive regimes including those beyond the Brownian limit. The GLE for a single particle in the absence of a net external force can be written as
\begin{equation}
\frac{dp(t)}{dt}=-\int_{0}^{t}\Pi(t-t')p(t')dt'+\eta(t)\text{,}\label{GLE}
\end{equation}
where $\Pi(t)$ is the memory function, and $\eta(t)$ is a random force of zero mean. Besides this, the noise is uncorrelated with the initial value $p(0)$, $\langle\eta(t)p(0)\rangle=0$, and obeys the fluctuation-dissipation theorem (FDT)~\cite{Kubo66}:
\[
\langle\eta(t)\eta(t')\rangle=\langle p^{2}\rangle_{eq}\Pi(t-t')\text{,}
\]
where $\langle\ldots\rangle_{eq}$ is an average over an ensemble in thermal equilibrium. The solution of the GLE is 
\begin{equation}
p(t)=p(0)R(t)+\int_{0}^{t}R(t-t')\eta(t')dt'\text{,}\label{p}
\end{equation}
where $R(t)$ can be obtained by multiplying Eq.~(\ref{GLE}) by $p(0)$ and taking the ensemble average, 
\begin{equation}
\frac{dR(t)}{dt}=-\int_{0}^{t}R(t')\Pi(t-t')dt'\text{,}\label{dR}
\end{equation}
whose Laplace transform yields $\tilde{R}(z)=1/[z+\tilde{\Pi}(z)]$. In this sense, $ \chi(t)\equiv -dR(t)/dt $ 
acts as a response function~\cite{Kubo66}. 

\emph{4. Ensemble average and equilibrium condition.}---If a system is ergodic and there are no external forces, thermal equilibrium should be observed in a time $t\gg\tau_{c}$, where $\tau_{c}$ is a relaxation time. Then, the distribution function of $p$ approaches the equilibrium distribution in the limit $t\rightarrow\infty$, and the mean energy converges to the equilibrium value, $\langle p^{2}(t\rightarrow\infty)\rangle=\langle p^{2}\rangle_{eq}$.

For any initial distribution of values, $p(0)$, it is possible to obtain the temporal evolution of the moments $\langle p^{n}(t)\rangle$, with $n=1,2,\ldots$. The first moment is obtained directly by taking the ensemble average of Eq.~(\ref{p}):
\begin{equation}
\langle p(t)\rangle=\langle p(0)\rangle R(t)\text{.}\label{p_ave}
\end{equation}
 Taking the square of Eq.~(\ref{p}) and performing an ensemble average, we get
\begin{equation}
\langle p^{2}(t)\rangle=\langle p^{2}\rangle_{eq}+R^2(t)\left[\langle p^{2}(0)\rangle-\langle p^{2}\rangle_{eq}\right]\text{.}\label{p2_ave}
\end{equation}
Consequently, we see that the knowledge of $R(t)$ allows one to describe completely these averages. 
  Equations (\ref{p_ave}) and (\ref{p2_ave}) are sufficient to show the condition of equilibrium for diffusion: if condition (\ref{R(t->infty)}) holds, then the time evolution will produce the ensemble average with $\langle p(t\rightarrow\infty)\rangle=0$ and $\langle p^{2}(t\rightarrow\infty)\rangle=\langle p^{2}\rangle_{eq}$. This result also suggests that the EH holds, and thus the KT holds. Now we may ask in what situation Eq.~(\ref{R(t->infty)}) is not valid. First, one should note that the long time behavior is associated with the small values of $z$ in the Laplace transform. Indeed, from the final value theorem~\cite{Gluskin03} we have
\begin{equation}
\lim_{t\rightarrow\infty}R(t)=\lim_{z\rightarrow0}z\tilde{R}(z)\text{.}\label{final}
\end{equation}
Therefore, it is only necessary to know $\tilde{R}(z)$. Morgado \textit{et al.}~\cite{Morgado02} obtained a general relationship between the Laplace transform of the memory function $\tilde{\Pi}(z)$ and the diffusion exponent $\alpha$:
\begin{equation}
\lim_{z\rightarrow0}\tilde{\Pi}(z)\approx cz^{\alpha-1}\text{,}\label{rafa}
\end{equation}
where $c$ is a positive non-dimensional constant. Now, using Eq.~(\ref{rafa}), in the limit Eq.~(\ref{final}) we have
\begin{equation}
\lim_{t\rightarrow\infty}R(t)=\lim_{z\rightarrow0}\left(1+cz^{\alpha-2}\right)^{-1}\text{,}
\label{memory}
\end{equation}
which is null for all diffusive processes in the range $0<\alpha<2$. In fact, this occurs in equilibrium or near-equilibrium states in which the validity of Linear Response Theory holds. On the other hand, this condition fails for ballistic motion, $\alpha=2$, in which $R(t\rightarrow\infty)=1/(1+c)$ and the autocorrelation function $C_{p}(t)$ will be non-null for long times. In other words, if the ballistic system is not initially equilibrated, then it will never reach equilibrium and the final result of any measurement will depend on the initial conditions. In this situation, the EH will not be valid; however, once again the KT holds since the violation of the EH was due to the violation of the irreversibility condition, Eq.~(\ref{R(t->infty)}), as predicted by Khinchin. The main consequence of the violation of this condition is the presence of a residual current, Eq.~(\ref{p_ave}). However, the effective current can be very small compared to $\langle p(0)\rangle$ and its value, as any other measurable property for ballistic diffusion, will depend on the value of $c$. In other words, the system decays to a metastable state and remains in it indefinitely, even in the absence of an external field.

\emph{5. Time average and the EH.}---The time averages of correlation functions are crucial for elucidating the properties of dynamical processes and play an extremely important role in the ergodic theory and, consequently, in physics. For diffusive systems governed by the GLE,  we will show that  the condition $R(t\rightarrow\infty)=0$ is sufficient for the time average to be equivalent to the ensemble average, \emph{i.e.}, for the system to be ergodic. For macroscopic systems with a large number of degrees of freedom, the effect of past values of the forces usually vanishes for a sufficiently large $t$, and the aforementioned condition is quite reasonable.

Let us consider the time average integral
\begin{equation}
I_{ta}=\lim_{T\rightarrow\infty}\frac{1}{T}\int_{0}^{T}\int_{0}^{t}\chi(t,t')dt'dt\text{.}\label{Ita}
\end{equation}
For stationary systems, $\chi(t,t')=\chi(t-t')$, we arrive at~\cite{Lee07a,Lee01}
\begin{equation}
I_{ta}=\lim_{t\rightarrow\infty}\left[\int_{0}^{t}\chi(t')dt'+R(t)-\frac{1}{t}\int_{0}^{t}R(t')dt'\right].\label{Ita2}
\end{equation}
Given that $R(t)$ is a real-valued function that converges asymptotically to a finite value, since we are working with the velocity autocorrelation, we can use a generalization of the final-value theorem for Laplace transforms~\cite{Gluskin03},
\[
\lim_{z\rightarrow0}z\tilde{R}(z)=\lim_{T\rightarrow\infty}\frac{1}{T}\int_{0}^{T}R(t)dt.
\]
With this, we obtain from Eq.~(\ref{Ita2})
\begin{equation}
\tilde{\chi}(0)+R(t\rightarrow\infty)-\lim_{z\rightarrow0}z\tilde{R}(z)=\chi_{s}\text{,}\label{tildeChi1}
\end{equation}
where $\chi_{s}$ is the time independent value, often called static susceptibility. On the other hand, taking the Laplace transform of Eq.~(\ref{dR}), we obtain $ \tilde{\chi}(z)+z\tilde{R}(z)=\chi_{s}$. Taking the limit $z\rightarrow0$, the previous relation becomes
\begin{equation}
\tilde{\chi}(0)+\lim_{z\rightarrow0}z\tilde{R}(z)=\chi_{s}\text{.}\label{tildeChi2}
\end{equation}
Comparing Eq.~(\ref{tildeChi2}) with Eq.~(\ref{tildeChi1}), one should conclude that the EH can only be valid if $R(t\rightarrow\infty)=0$, \emph{i.e.} if the irreversibility condition (\ref{R(t->infty)}) holds. From Eq.~(\ref{final}) we end up with
\begin{equation}
\tilde{\chi}(0) = \chi_{s}.
\end{equation}
 Again this is a consequence of  the irreversibility condition. Therefore, irreversibility is a necessary and sufficient condition for the EH to hold in diffusive processes described by a GLE.
\begin{figure}[!h]
\includegraphics[scale=0.45]{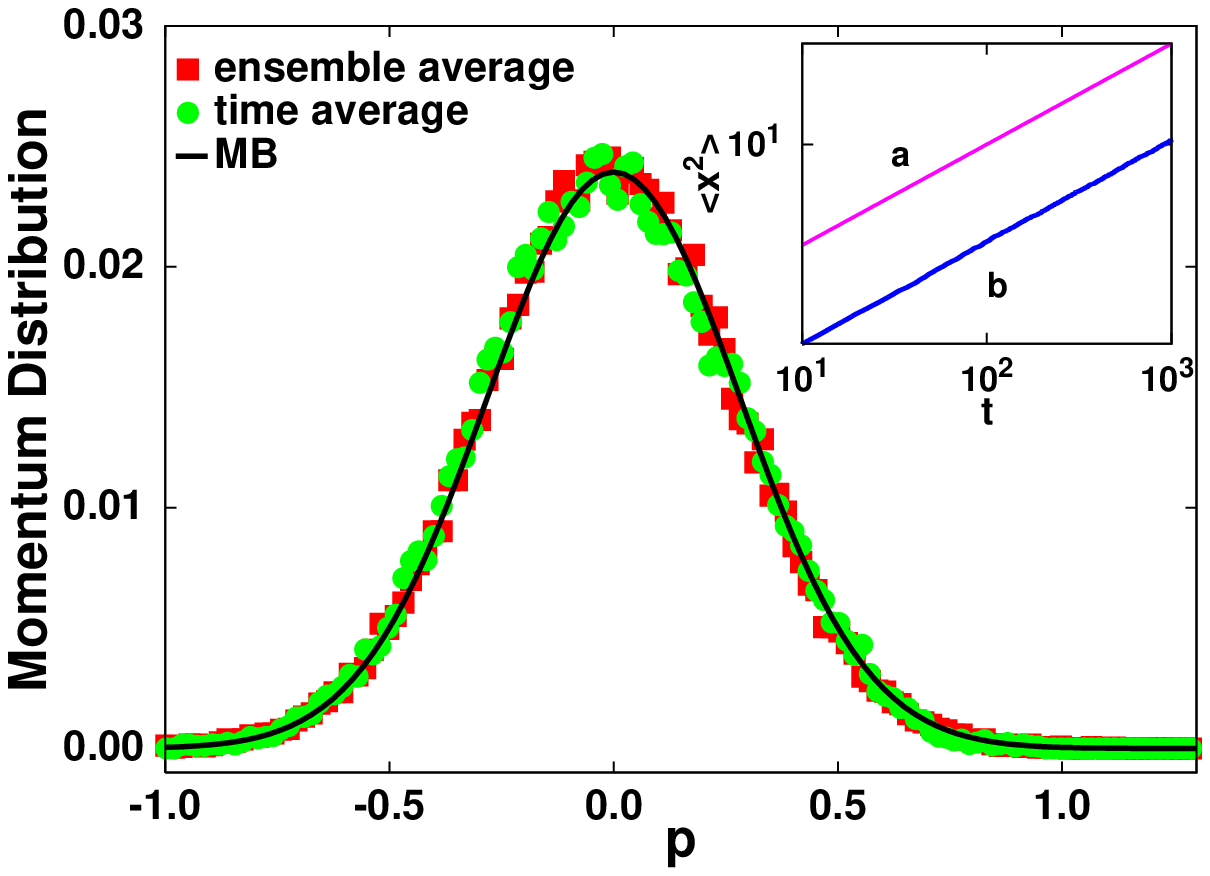}
\includegraphics[scale=0.45]{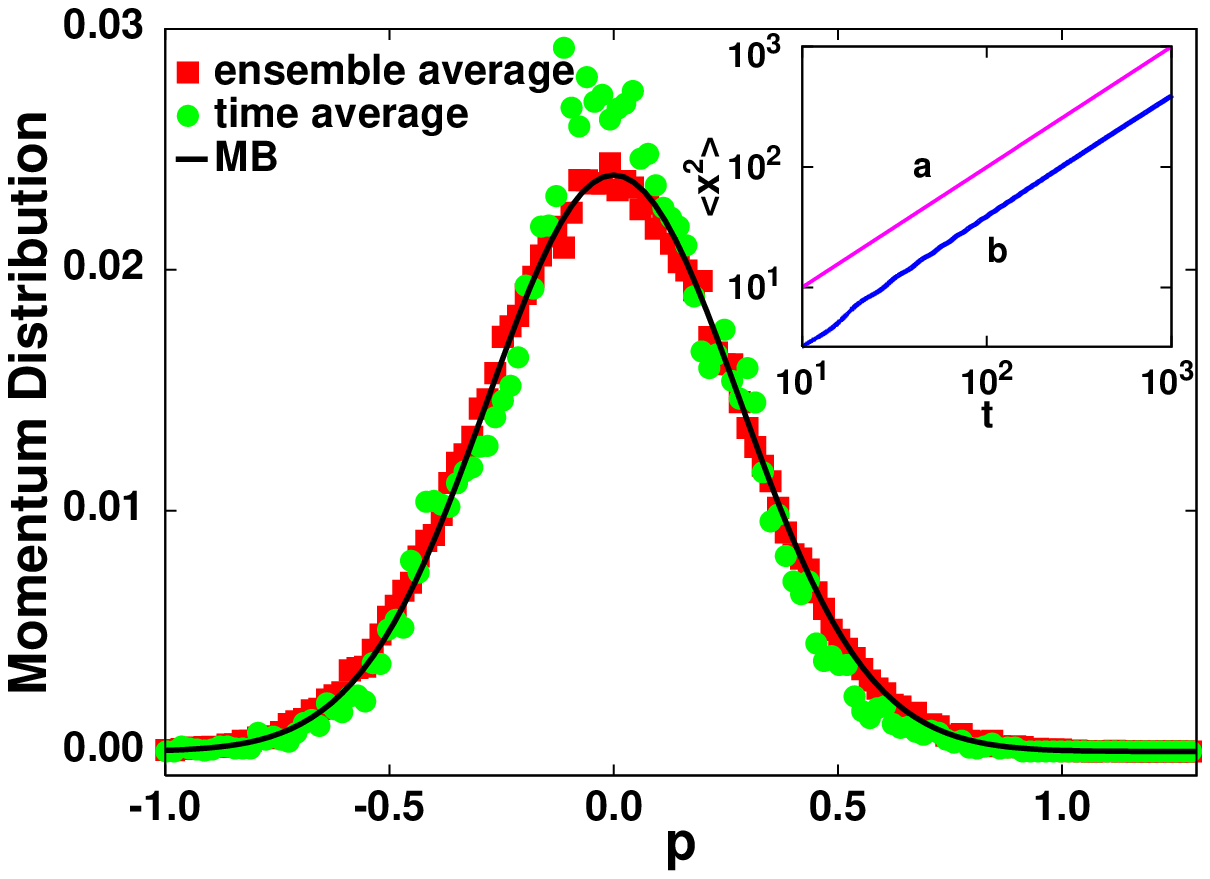}
\includegraphics[scale=0.45]{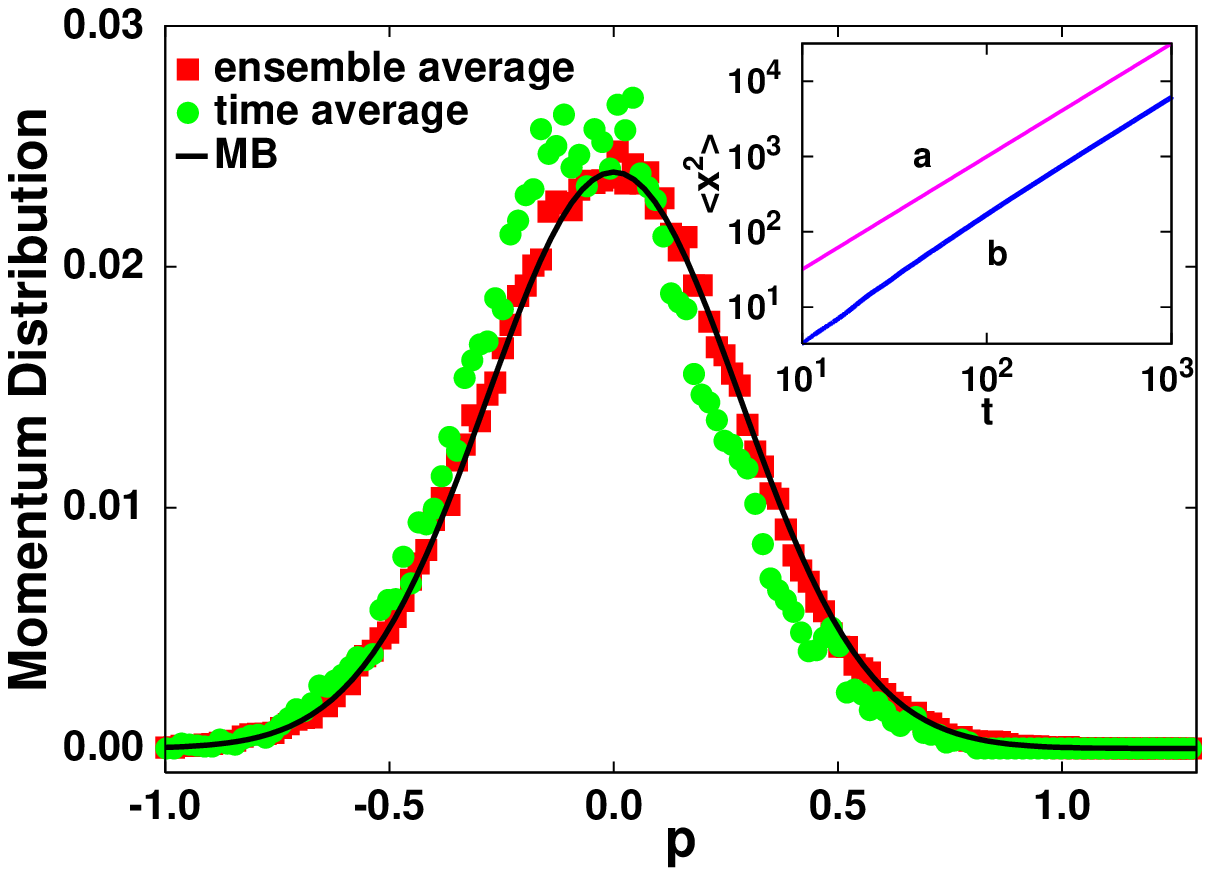}
\caption{(Color online) Numerical results for the probability distribution function for subdiffusion (top, $\alpha=0.5$), normal diffusion (middle, $\alpha=1$) and superdiffusion (bottom, $\alpha=1.5$). The time averages (circles) are obtained by following one particle trajectory and calculating the histogram for times from $t=100$ to $t=5000$. For the ensemble averages (squares), we calculate the histogram using $5\cdot10^{4}$ particles, at time $t=1000$. The continuous line is the Maxwell-Boltzmann distribution. Insets: Curves $a$ correspond to the functions $t^\alpha$ and curves $b$ to the simulated mean square displacements. }
\label{figure}
\end{figure}

\emph{6. Simulation.}--- In order to illustrate the analytical results, we have numerically integrated the GLE, Eq.~(\ref{GLE}), to obtain approximations to the probability distribution of particle velocities using histograms. We construct the memory using
\begin{equation}
\Pi(t)=\int \rho(\omega) \cos (\omega t ) d\omega\text{,} \label{Pi}
\end{equation}
 where
\begin{equation}
\rho(\omega) = \left\{
\begin{array}{cc}
a \omega^{\beta}, &\text{ for } \omega\leq\omega_{s}\\
g(\omega), & \text{ otherwise.}
\end{array}\right.
\end{equation}
 Here, the function $g(\omega)$ is arbitrary as long as it is sufficiently well-behaved and that its integral in the memory function converges. If one is interested only in the long time behavior $t\gg 1/\omega_s$, it can be taken to be $0$. With this noise density of states, it is possible to simulate many diffusive regimes~\cite{Vainstein06a}. Noise of this form can be obtained either by formal methods or empirical data. Using this expression in Eq.~(\ref{Pi}), taking the Laplace transform in the limit $z \rightarrow 0$, we have 
\begin{equation}
\tilde{\Pi}(z) \propto \left\{
\begin{array}{cc}
z^{\beta}, &\text{ for } \beta<1,\\
-az\ln(z), &\text{ for } \beta=1,\\
z, & \text{ for } \beta>1.
\end{array}\right.
\end{equation}
Consequently, for this type of noise, there is a maximum value of $\alpha$, \emph{i.e.}, $\alpha\leq 2$ for any value of $\beta$. It should be noted that the case $\beta=1$ does not lead to a memory whose Laplace transform is in the form of Eq.~(\ref{rafa}).  For $-1<\beta<1$, we obtain $\alpha = 1 + \beta$. For $\beta>1$, one has $\alpha = 2$, which shows that ballistic diffusion is a limiting case for the GLE with this type of memory.

 In Fig.~(\ref{figure}) we show the probability distribution functions obtained for subdiffusion ($\beta=-0.5$), normal diffusion ($\beta=0$), and superdiffusion ($\beta=0.5$) for the values $a=0.25$ and $g(w)=0$. We have used $\omega_s=0.5$ for all cases except for subdiffusion, which demands a broad noise $\omega_s=2$ to reach the stationary state. In all cases, we expect that $R(t\rightarrow\infty)=0$, and that the EH will be valid even for the subdiffusive (superdiffusive) case, despite the fact that $W=0$ ($W=\infty$). These relations can be seen by considering the limit $W = \lim_{z\rightarrow 0} \tilde{R}(z)$. If the EH is valid, the velocity probability distribution will be the same for an average over an ensemble of particles and for a time average over the trajectory of a single particle for long times after the system has reached an equilibrium state. Note that despite the presence of large fluctuations in the time average case due to numerical errors, there is a good agreement between the resulting ensemble and time distributions. The three probability distributions converge toward the Maxwell-Boltzmann distribution,  which is in accordance with previous analytical results~\cite{Lapas07}. 

\emph{7. Concluding remarks.}---In this work, we have shown that the KT (proved by Khinchin for normal diffusion) holds for all kinds of diffusive processes, which are ergodic in the range of exponents $0<\alpha<2$. This result may have deep consequences in many areas~\cite{Ediger96,Kauzmann48,Parisi97,Ricci00,Margolin05,Hentschel07}. Moreover, it could be verified in and applied to experimental systems, such as the subdiffusive dynamics of the distance between an electron transfer donor and acceptor pair within a single protein molecule~\cite{Yang03}, which has recently been modelled by a GLE~\cite{Kou04_and_Dua08}. Such a model successfully explains the equilibrium fluctuations and its broad range of time scales, being in excellent agreement with experiments. The KT gives the EH a practical character, since it is expressed in terms of response functions: our results apply for real-valued relaxation functions $R(t)$; on the other hand, if the relaxation function assumes complex values, e.g. conductivity,  the final value theorem may not be applied. For those systems, the KT fails, as proposed in Ref.~\cite{Lee07a}.

In principle, it is generally possible to derive a GLE for Markovian systems by eliminating variables, whose effects are incorporated in the memory kernel and in the colored noise~\cite{Helleman80}. Altogether, some results obtained for the GLE formalism should be valid for diffusion described by fractional Fokker-Planck equations, since both formalisms yield similar results~\cite{Kou04_and_Dua08}. For nonlinear Hamiltonian maps~\cite{Shlesinger93} there is no general framework to address the problem and, in particular, the absence of a  coupling to a thermal bath (explicit in the GLE) and consequently the lack of a detailed balance relation or FDT may require a specific analysis of each case. However, since it is possible to give a kinetic description of the Hamiltonian dynamics by means of a fractional Fokker-Planck-Kolmogorov equation~\cite{Zaslavsky02}, it is expected that the treatment of anomalous diffusion in such systems should also be possible by the GLE formalism. Further research in this direction is needed and will open new perspectives.

\acknowledgments  This work was supported by Brazilian Research Foundations: CAPES, CNPq, FINATEC, and FAPDF; and by the DGiCYT of the Spanish Government under Grant No. FIS2005-01299.

\bibliographystyle{unsrt}

\end{document}